\documentclass{emulateapj}

\newcommand{\mic}{\,$\mu$m }
\newcommand{\micpa}{\,$\mu$m}          
\newcommand{\muJy}{\,$\mu$Jy }

\newcommand{\Lsol}{L$_\odot$}
\newcommand{\Msol}{M$_\odot$}

\newcommand{\cxo}{CXO--J1417 }
\newcommand{\cxopa}{CXO--J1417}
\bibpunct[]{(}{)}{;}{a}{}{,}
 
\shorttitle{On the coeval growth of bulges and massive black holes}
\shortauthors{E.\,Le Floc'h et al.}

\begin{document}
\def\gtapp
{\mathrel{\hbox{\raise0.3ex\hbox{$>$}\kern-0.8em\lower0.8ex\hbox{$\sim$}}}}
\def\ltapp
{\mathrel{\hbox{\raise0.3ex\hbox{$<$}\kern-0.75em\lower0.8ex\hbox{$\sim$}}}}
\def\ts{\thinspace}

\title{Far-infrared characterization of an ultra-luminous starburst associated
with a massively-accreting black hole at $z$\,=\,1.15}

\slugcomment{Accepted for publication in the Astrophysical Journal Letters, 
September 15th, 2006}

\author{E.~Le~Floc'h\altaffilmark{1,2},
C.N.A.\,Willmer\altaffilmark{1}, 
K.\,Noeske\altaffilmark{3},
N.P.\,Konidaris\altaffilmark{3}, 
E.S.\,Laird\altaffilmark{3}
D.C.\,Koo\altaffilmark{3},
K.\,Nandra\altaffilmark{4},
K.\,Bundy\altaffilmark{5},
S.\,Salim\altaffilmark{6},
R.\,Maiolino\altaffilmark{7},
C.J.\,Conselice\altaffilmark{8},
J.M.\,Lotz\altaffilmark{9},
C.\,Papovich\altaffilmark{1,10},
J.D.\,Smith\altaffilmark{1},
L.\,Bai\altaffilmark{1},
A.L.\,Coil\altaffilmark{1,10},
P.\,Barmby\altaffilmark{11},
M.L.N.\,Ashby\altaffilmark{11},
J.-S.\,Huang\altaffilmark{11},
M.\,Blaylock\altaffilmark{1},
G.\,Rieke\altaffilmark{1},
J.A.\,Newman\altaffilmark{12,10},
R.\,Ivison\altaffilmark{13},
S.\,Chapman\altaffilmark{5},
H.\,Dole\altaffilmark{14},
E.\,Egami\altaffilmark{1} \&
D.\,Elbaz\altaffilmark{15} 
}

\altaffiltext{1}{Univ. of Arizona, Tucson, AZ 85721; elefloch, cnaw,
    papovich, jdsmith, bail, acoil, blaylock, grieke, eegami@as.arizona.edu} 

\altaffiltext{2}{Chercheur Associ\'e, Observatoire de Meudon, Paris,
  France}

\altaffiltext{3}{UCO/Lick Observatory, Univ. of California, Santa Cruz, CA
  95064; kai, npk, eslaird, koo@ucolick.org}

\altaffiltext{4}{Astrophysics group, Imperial College, London SW7 2AW,
  UK; k.nandra@imperial.ac.uk}

\altaffiltext{5}{California Institute of Technology, Pasadena, CA 91125;
kbundy, schapman@astro.caltech.edu}

\altaffiltext{6}{UCLA,
 Los Angeles CA 90095; samir@astro.ucla.edu}

\altaffiltext{7}{INAF - Rome I-00044,
  Italy;  maiolino@oa-roma.inaf.it}

\altaffiltext{8}{University of
  Nottingham, UK; conselice@nottingham.ac.uk}

\altaffiltext{9}{NOAO Leo Goldberg Fellow, Tucson, AZ 85719;
  lotz@noao.edu}

\altaffiltext{10}{Spitzer/Hubble Fellows}

\altaffiltext{11}{CfA,
  Cambridge, MA 02138; pbarmby, mashby, willner,
  jhuang@cfa.harvard.edu}

\altaffiltext{12}{Lawrence Berkeley Nat. Lab.,
 CA 94720; janewman@lbl.gov}

\altaffiltext{13}{Royal Observatory, Edinburgh EH9 3HJ, UK; rji@roe.ac.uk}

\altaffiltext{14}{IAS, F-91405 Orsay,
  France; herve.dole@ias.u-psud.fr}

\altaffiltext{15}{SAp-CEA, 91191
  Gif-sur-Yvette, France; delbaz@cea.fr}

\begin{abstract} 
  As part of the {\it All Wavelength Extended Groth Strip
    International Survey (AEGIS),\,} we describe the panchromatic
  characterization of an X-ray luminous active galactic nucleus (AGN)
  in a merging galaxy at $z$\,=\,1.15.  This object is detected at
  infrared (8\micpa, 24\micpa, 70\micpa, 160\micpa), submillimeter
  (850\micpa) and radio wavelengths, from which we derive a bolometric
  luminosity L$_{\rm bol}$\,$\sim$\,9$\times$10$^{12}$\,\Lsol.  We
  find that the AGN clearly dominates the hot dust emission below
  40\mic but its total energetic power inferred from the hard X-rays
  is substantially less than the bolometric output of the system.
  About 50\% of the infrared luminosity is indeed produced by a cold
  dust component that probably originates from enshrouded star
  formation in the host galaxy.  In the context of a coeval growth of
  stellar bulges and massive black holes, this source might represent
  a ``transition'' object sharing properties with both quasars and
  luminous starbursts.  Study of such composite galaxies will help
  address how the star formation and disk-accretion phenomena may have
  regulated each other at high redshift and how this coordination may
  have participated to the build-up of the relationship observed
  locally between the masses of black holes and stellar spheroids.
\end{abstract}


\section{Introduction}

Galaxies with a bolometric luminosity exceeding 10$^{12}$\,\Lsol \,
are often powered by a combination of massive star formation and
accretion of material around active nuclei \citep[e.g.,][]{Genzel98}.
Early in cosmic history the connection between these two phenomena may
have led to a coeval growth of super massive black holes (SMBHs) and
stellar spheroids \citep[e.g.,][]{Page01}, and could thus be the
foundation of the correlation observed between the masses of these two
components in the local Universe \citep[e.g.,][]{Gebhardt00}.
However, the implication of such a co-evolution in the more general
context of the stellar mass built-up history
\citep[e.g.,][]{Dickinson03} and SMBH formation
\citep[e.g.,][]{Barger05} has not been fully addressed.  Understanding
the importance of this ``coordinated'' activity of starbursts and
active galactic nuclei (AGNs) requires properly deconvolving their
respective contributions {\it within individual objects,} a
challenging task particularly for distant sources.

Because of the complexity of the AGN and starburst spectral energy
distributions (SEDs), this decomposition can best be achieved by
combining data from wavebands that offer distinctive spectral features
to characterize these phenomena.  For instance, active nuclei are
strong emitters at high energy and they are usually associated with a
continuum of hot dust peaking in the mid-infrared (IR). Luminous
starbursts, on the other hand, are characterized by a colder dust
component and emit the bulk of their luminosity in the far-IR.  To
illustrate how the coexistence of these two processes at high redshift
could be studied using larger samples, we present a panchromatic
analysis of a luminous AGN referenced as CXO--GWS--J141741.9+522823
(hereafter CXO-J1417) by \citet{Nandra05}. It is embedded in an
ultra-luminous infrared galaxy (ULIRG) at $z$\,=\,1.15 and its
detection across the full electromagnetic spectrum allows us to
constrain the level of star formation also present in this source.  We
assume a $\Lambda$CDM cosmology with
H$_0$\,=\,70~km~s$^{-1}$\,Mpc$^{-1}$, $\Omega_m$\,=\,0.3 and
$\Omega_{\lambda}\,=\,0.7$.

\section{The data}

CXO-J1417 is a bright X-ray source also known in the literature as
CFRS\,14.1157 or CUDSS 14.13.  It is associated with a very red galaxy
($I$--$K_{AB}$\,$\sim$\,2.6: \citealt{Webb03b}; see also
\citealt{Wilson06}) detected in the mid-IR
\citep{Flores99,Higdon04,Barmby06,Ashby06}, submillimeter
\citep{Eales00,Webb03b} and radio \citep{Fomalont91,Chapman05}. High
resolution images obtained with {\it HST\,} \citep{Davis06,Lotz06}
reveal several components presumably interacting with each other (see
Fig.\,1).  The brightest is dominated by a point source located at
RA\,=\,14$^h$17$^m$41\,\fs89 and Dec\,=\,52$^{\rm
  o}$28\arcmin23.65\arcsec (J2000,
$\delta$RA\,$\sim$\,$\delta$Dec\,$\sim$\,0.07\arcsec) coinciding
precisely with the position of the X-ray detection \citep{Miyaji04}.

\begin{figure}[htpb]
  \epsscale{1.1}
\plotone{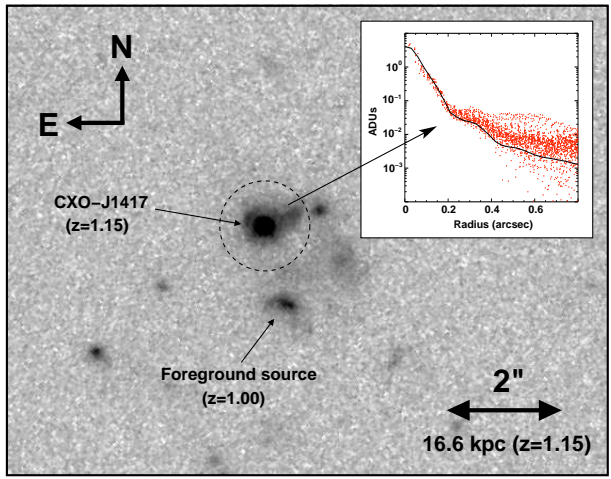}
\caption{An F814W ACS image of \cxo ($z$\,=\,1.15) with the 95\% error location of the X-ray source indicated
by the dashed-line circle \citep{Miyaji04}.
 The inset illustrates the profile of this component in the F814W image.
The comparison with the 
PSF (solid line) reveals that the central core is not resolved by $HST$.
The object lying 1.4\arcsec \,
to the South 
 is a foreground galaxy at $z$\,=\,1.00.}
\end{figure}

At this location
there is also a bright and point-like object detected at 70\mic and
160\micpa.
 In spite of the large
beam used for the data at long wavelengths\footnote{The FWHM of the
  MIPS PSF is 18\arcsec \, and 40\arcsec \, at 70\mic and 160\mic
  respectively.}, the fact that the mid-IR counterpart of \cxo is two
orders of magnitude brighter than any other galaxies detected at
24\mic in this area strongly suggests that this far-IR emission is also
associated with the X-ray source.
Its fluxes in the MIPS 24/70/160\mic bands were
  measured
via PSF fitting. They are reported in Table\,1, which also summarizes
the full multi-wavelength photometry of the object (see
\citealt{Davis06} for a description of our data set).  Optical and
near-IR fluxes were measured within a 1\arcsec-radius aperture
centered at the position of \cxopa.
With the exception of the GALEX data where we believe that the emission is
 contaminated by a   blue galaxy 
lying 1.4\arcsec \, to the South, 
 our flux measurements   refer exclusively to
the component hosting the X-ray source.

\begin{deluxetable}{lcc}
\setlength{\tabcolsep}{0.4in}
\tablecaption{Photometry of \cxo
\label{tab:conf}}
\tablehead{
\colhead{Band} &
\colhead{Flux/Flux density$^{a}$} &
\colhead{Reference$^{b}$} 
}
\startdata
2--10\,keV      & 3.8$\pm$0.3$\times$10$^{-14}$\,erg\,cm$^{-2}$\,s$^{-1}$   & 1         \\
0.5--2\,keV     & 1.3$\pm$0.1$\times$10$^{-14}$\,erg\,cm$^{-2}$\,s$^{-1}$   & 1         \\
FUV (1539\AA)   & $<$0.45\muJy (3$\sigma$)                               &  \\
NUV (2316\AA)   & 0.93$\pm$0.15\muJy                                     &\\
$B$ (4389\AA)   & 0.8$\pm$0.1\muJy                                       &         \\
$R$ (6601\AA)   & 6.05$\pm$0.15\muJy                                     &          \\
$I$ (8133\AA)   & 16.7$\pm$0.8\muJy                                      &         \\
$J$ (1.2\micpa) & 57.6$\pm$0.5\muJy                                      &          \\
$K$ (2.2\micpa) & 117$\pm$1\muJy                                         &         \\
IRAC 3.6\mic    & 580.1$\pm$0.4\muJy                                     &  2, 3        \\
IRAC 4.5\mic    & 981.7$\pm$0.5\muJy                                     &  2, 3        \\
IRAC 5.8\mic    & 1448$\pm$4\muJy                                        &  2, 3        \\
IRAC 8.0\mic    & 2225$\pm$4\muJy                                        &  2, 3        \\ 
IRS 16\mic      & 3.3$\pm$0.7\,mJy                                       &  4        \\
MIPS 24\mic     & 5.75$\pm$\,0.1mJy                                      &  \\
MIPS 70\mic     & 20.1$\pm$1.2\,mJy                                      &  \\
MIPS 160\mic    & 105$\pm$30\,mJy                                        & \\
SCUBA 850\mic   & 3.3$\pm$1\,mJy                                         & 5         \\
VLA 5\,GHz      & 53.6$\pm$4\muJy                                        & 6         \\
VLA 1.4\,GHz    & 110$\pm$40\muJy                                        & ~ 6, 7, 8, 9  
\enddata
\vspace{-.0cm}
\tablenotetext{a}{\,Optical/near-IR fluxes are measured in a 1\arcsec-radius aperture.}
\tablenotetext{b}{\,References -- 1: \citet{Nandra05}; 
2: \citet{Barmby06};
3: \citet{Ashby06}; 
4: \citet{Higdon04}; 
5: \citet{Eales00};
6: \citet{Fomalont91}; 
7: \citet{Webb03b};
8: \citet{Chapman05};
9: \citet{Ivison06}. }
\end{deluxetable}

A redshift of $z$\,=\,1.15 was reported by \citet{Hammer95} based on
UV/optical spectroscopy.  In Fig.\,2 we display the combined spectrum
obtained at Keck by \citet{Davis03} using LRIS and DEIMOS.  The
presence of an AGN is clearly confirmed by the detection of [NeV] and
MgII.  There is however considerable self-absorption of the latter
\citep{Sarajedini06}, which prevents firm classification as a type~1
or a type~2 object. Interestingly, we also note the detection of
Ca\,K$+$H absorption lines redshifted by
$\sim$\,150--200\,km\,s$^{-1}$ relative to [OII] and which suggests
the presence of gas inflow in the galaxy.

\begin{figure}[htpb]
\epsscale{1.1}
\plotone{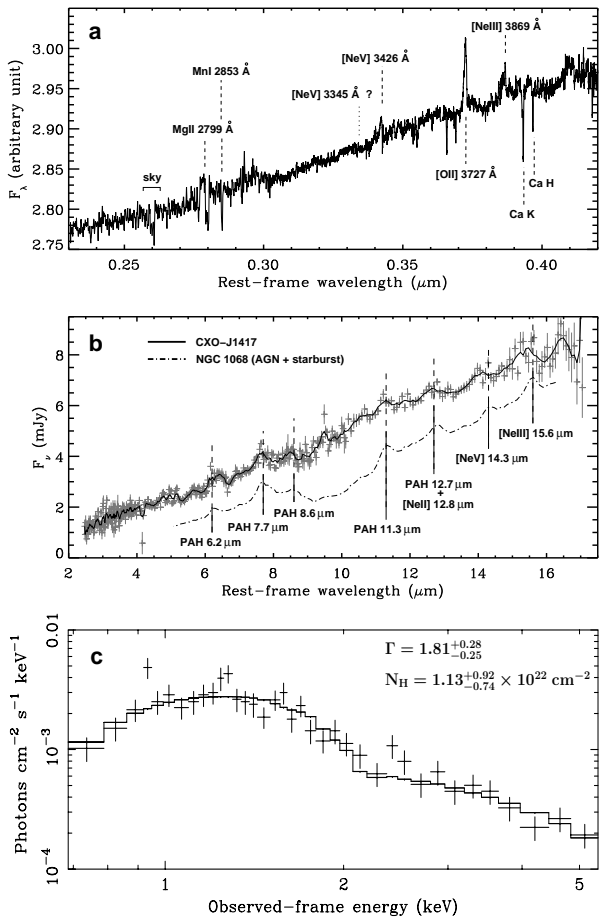}
\caption{The optical ({\it a\,}), mid-IR ({\it b\,}) and X-ray ({\it
    c\,}) spectra of \cxopa. The X-ray properties and the detection of
  [NeV]\,3426\AA \, reveal the presence of an AGN that is also
  responsible for the hot dust emission and the feature-less power-law
  SED in the mid-IR.  The solid-line in panel {\it b\,} is a smoothed
  version of the observed spectrum, while the dashed-dotted line shows
  a comparison with the mid-IR SED of the Seyfert\,2 NGC\,1068.  The
  solid-line in panel {\it c\,} represents the best fit to the data,
  obtained with the parameters mentioned in the top-right corner.  In
  spite of the lack of strong silicate absorption at 9.7\micpa, the
  column density derived at high energy indicates a significant
  obscuration toward the nucleus.  }
\end{figure}

Finally, mid-IR spectroscopy was carried out by \citet{Higdon04} as
well as our own group.  We obtained\footnote{General Observer program
  ID 3\,216.}  960\,s total exposure time for each of the two Short
Low (SL1: 7.4--14.5\micpa; SL2:5.2--8.7\micpa) and each of the two
Long Low modules of the IRS (LL1: 19.5--38\micpa; LL2:
14.0--21.3\micpa, see \citealt{Houck04a}), while \citet{Higdon04}
targeted \cxo for a total integration of 1440\,s in LL1 and LL2.  We
reduced and combined all these data using the pipeline developed by
the FEPS Legacy team \citep{Hines06}, thus bringing the total
integration to 2400\,s for both LL1 and LL2.  Our final spectrum
covers the rest-frame 2.5--16.5\mic range. It is displayed in Fig.\,2
along with the X-ray spectrum of \cxo obtained by \citet{Nandra05}.

\section{A panchromatic characterization of \cxopa}

The full spectral energy distribution of \cxo is illustrated in
Fig.\,3.  It has a very red continuum with a steep rise from the
optical up to the mid-IR, and the rest-frame 1.6\mic ``bump'' usually
associated with stellar populations \citep{Sawicki02} is not detected
\citep{Barmby06,Ashby06}.  The lack of this stellar feature and the
very strong hot-dust power-law emission that we observe at
$\sim$1--5\mic \citep[see also][]{Higdon04} are characteristic of
active nuclei \citep[e.g.,][]{Brand06}. They reveal that the AGN
totally dominates the luminosity of \cxo at these short IR
wavelengths.

\begin{figure}[htpb]
\epsscale{1.1}
\plotone{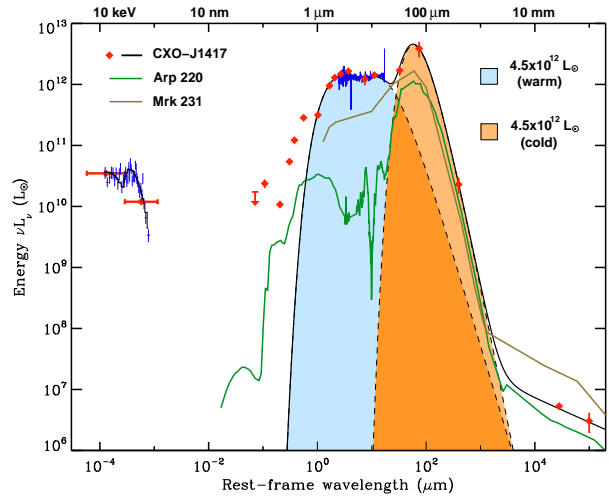}
\caption{ Panchromatic  SED of \cxo (red diamonds)
overlaid with a fit covering the X-ray, IR and radio wavelength range
(solid black line, see text for details). The X-ray and mid-IR spectra
from {\it Chandra\,} and IRS are also displayed (blue symbols and blue
line respectively).  The IR portion is fitted by a two-component
model accounting for the warm (light-blue shaded) and cold (light-red
shaded) dust emission. The fit in the radio assumes synchrotron
emission with a spectral index $\alpha$\,=\,0.6.  The SEDs of Arp\,220
and Mrk\,231 (adapted from \citealt{Silva04}, \citealt{Spoon04} and \citealt{Ivison04})
are shown for comparison
(green and brown lines respectively).} 
\end{figure}

We explored in more detail the properties of this active nucleus by
fitting the X-ray spectrum with a power-law intrinsically absorbed at
$z$\,=\,1.15 and assuming a Galactic extinction model.  We derived a
column density N$_{\rm
  H}$\,=\,1.13$^{+0.92}_{-0.74}$\,$\times$\,10$^{22}$\,cm$^{-2}$ and a
photon index $\Gamma$\,=\,1.81$^{+0.28}_{-0.25}$, leading to an
extinction-corrected luminosity L$_{\rm
  2-10\,keV}$\,=\,2.35$^{+0.30}_{-0.31}$\,$\times$\,10$^{44}$\,erg\,s$^{-1}$.
Although the lack of significant extinction by the silicates at
9.7\mic and the non-detection of the 400\,eV iron K$\alpha$ line
indicate that it is not an extremely absorbed object, the X-ray data
point therefore to a luminous AGN characterized by substantial
obscuration.

This active nucleus is probably not the only source powering the
bolometric luminosity of \cxopa. Although they are strongly diluted by
the continuum emission from the AGN, the mid-IR broad bands from the
Polycyclic Aromatic Hydrocarbons (PAHs) often seen in star-forming
environments seem to be detected in our IRS spectrum (see Fig.\,2b).
Assuming a power-law continuum superimposed with a typical PAH
template and leaving the redshift of the latter as a free parameter,
we generated a series of simulated spectra that we compared to our
data. The $\chi^2$ shows a clear minimum when the PAH component is
shifted to the distance of \cxopa, which suggests that these features
are detected with relatively good confidence\footnote{At the redshift
  of \cxo $z_0$\,=\,1.15, the $\chi^2$ is reduced by a factor of
  $\sim$\,2 with respect to its median value measured over a redshift
  range $z_0\pm\Delta z$ with $\Delta z \sim 0.5$.}.  Furthermore, the
far-infrared and submillimeter detections reveal a very luminous cold
dust component typical of those observed in dusty star-forming
galaxies (see Sect.\,4).  To quantify its contribution relative to the
much warmer dust seen in the mid-IR, we decomposed the global IR SED
beyond 1\mic into (i) a single modified black body\footnote{We adopt
  the form $B_{\nu}(T) \times \nu^{\beta}$ where $T$ is the dust
  temperature and $\beta$ the dust emissitivity.}  accounting for the
cold component, characterized by a temperature $T$\,$\sim$\,40\,K and
a dust emissivity $\beta$\,$\sim$\,1.7, and (ii) a combination of
several blackbodies with temperatures ranging from 150\,K to~2000\,K
and reproducing the warm dust emission (see Fig.\,3).  We derived
1--1000\mic integrated luminosities of 4.5$\times$10$^{12}$\,\Lsol\,
for each of these two components. This leads to a total IR luminosity
L$_{1-1000\mu m}$\,=\,9.0$\pm$0.4\,$\times$\,10$^{12}$\,\Lsol, where
the uncertainty is driven by the determination of the temperature and
the emissivity of the cold dust emission.

\cxo is also remarkably quiet in the radio.  The spectral index
($\alpha$\,$\sim$\,0.6) is typical of synchrotron emission from
starbursts, though we cannot exclude a flatter continuum because of
the substantial uncertainty on the observed flux density at
1.4\,GHz. Assuming the standard far-IR/radio correlation
\citep{Condon92}, we would infer an IR luminosity $\sim$2.5$\times$
lower than that implied by our fit of the cold dust component.  As it
has already been observed at low redshift
\citep[e.g.,][]{Rieke80,Clemens04,Gallimore04} this radio faintness
could result from substantial free-free absorption in the interstellar
medium of the galaxy. It could also be due to a synchrotron deficiency
characteristic of a very recent episode of star formation
\citep{Roussel03}.

\section{Discussion}
\subsection{On the nature of \cxopa}

Although the bolometric correction for luminous and strongly-absorbed
X-ray sources has not been very well constrained so far, the
obscuration toward \cxo is still reasonable enough to allow a fairly
secure estimate of the total luminosity of the active nucleus in this
object.  Comparison of the SED shown in Fig.\,3 with typical AGN
templates \citep[e.g.,][]{Elvis94,Silva04} indicate that this
bolometric luminosity should typically range between
1$\times$10$^{12}$\,\Lsol \, and 3$\times$10$^{12}$\,\Lsol.  While the
AGN can thus power most of the hot dust detected in the system, it is
not energetic enough to account also for the far-IR emission.  We
argue that the cold dust component is produced by a deeply enshrouded
starburst in the host galaxy (see also e.g., \citealt{Waskett03}).
\cxo could be therefore a high redshift analog of some nearby
``composite'' ULIRGs where the contribution of the AGN to the
bolometric output is comparable to that of the star-forming activity
\citep{Farrah03}.  It could also be similar to other distant X-ray
selected sources that were detected at long wavelengths
\citep[e.g.,][]{Page01}, though this far-IR emission has been
sometimes assumed to originate from the active nucleus rather than
star formation \citep{Barger05}.

Assuming the calibration from \citet{Kennicutt98}, the IR luminosity
of the cold component translates into a star-formation rate
$SFR$\,$\sim$\,750\,\Msol\,yr$^{-1}$. Such enhanced levels of activity
usually occur within embedded and very compact regions surrounding the
cores of galaxies ($\sim$\,100--300\,pcs, \citealt{Soifer00}). It is
consistent with the absence of direct star formation signatures as
inferred from our UV/optical photometry, as well as from the ACS image
taking into account the spatial resolution of the $HST$ data (i.e.,
$\sim$\,1\,kpc at $z$\,=\,1.15).

\subsection{Implications}
Assuming a typical accretion efficiency $\epsilon$\,=\,0.1
\citep{Marconi04}, the luminosity of the AGN in \cxo translates into a
mass accretion rate d$M$/d$t$\,$\sim$\,3.1\,\Msol\,yr$^{-1}$. This is
typical of quasars at $z$\,$\sim$\,1 \citep{McLure04}, but it is
larger than the rates measured in sources experiencing similar levels
of starburst activity such as the more distant SCUBA sources
\citep{Alexander05}.  Furthermore, the bolometric luminosity of this
object and the obscuration toward its nucleus suggest that the gas
fueling and the accretion are occurring quite efficiently, probably
close to the Eddington limit. Under this hypothesis we would derive a
black hole mass of $\sim$\,1.4\,$\times$\,10$^{8}$\,\Msol. This is
typically an order of magnitude larger than the mass of the SMBHs
determined in the submillimeter galaxies, and it would be even larger
in the case of a sub-Eddington accretion.  These properties suggest
that \cxo is an object sharing characteristics with both
starburst-dominated galaxies and quasars, where violent star formation
is still happening while a massive black hole has already formed.
 
High redshift ULIRGs showing a mixture of star formation and AGN such
as \cxo could be interesting as tests of the evolutionary sequences
that have been proposed to understand the connection between the two
phenomena \citep[e.g.,][]{Sanders88a}. In such scenarios for instance,
merging galaxies first trigger powerful star formation, and as
material settles into the cores of these objects it feeds a
supermassive black hole that eventually emerges as a luminous AGN.
The latter can then produce strong winds and outflows that feed energy
back into the surrounding galaxy and may either quench or reactivate
star formation (\citealt{Springel05,Hopkins05,Silk05}).  In the case
of \cxo there is a dominant contribution of the nucleus in the near-IR
and it is not clear whether an underlying bulge has already formed in
the host galaxy. However, sources experiencing star formation and
disk-accretion that both radiate a similar amount of energy throughout
their lifetime would evolve toward massive galaxies that lie
significantly out of the local ``$M_{\rm BH} - \sigma$'' relationship
\citep{Page01}. Although the starburst and the AGN in \cxo are
characterized by roughly equal luminosities now, we infer that if
these two phenomena evolve together they may occur on quite different
time scales and regulate each other efficiently for the bulges and
SMBHs to grow in a coordinated manner.

Such transitional cases might be rare locally
\citep[e.g.,][]{Genzel98}. At higher redshift however, their
importance relative to the infrared/submillimeter or X-ray selected
objects where one type of activity (i.e., star formation or accretion)
largely dominates is not yet known. Interestingly, \cxo lies at the
knee of the 2--8\,keV luminosity function derived by \citet{Barger05}
at 0.8\,$\leq$\,$z$\,$\leq$\,1.2 but it is much more IR-luminous than
most of star-forming galaxies at this epoch of cosmic history
\citep{LeFloch05}.  Searching for similar objects at higher redshifts
when ULIRGs were a major component of the starbursting population
\citep{Blain02} should allow us to explore in more detail the role
that this coexistence of AGNs and starbursts within galaxies played in
shaping the present-day Universe.  Even though their identification
could be challenging, \cxo points to the type of evidence required for
this goal. Large data sets from existing surveys like {\it AEGIS\,}
should provide this information for enough sources to probe the
prevalence of this phase of galaxy evolution at $z$\,$\gtapp$\,1.

\acknowledgments We thank Roberto Gilli for his valuable comments as
well as Jim Cadien, Dean Hines and Jeroen Bouwman for their help in
the data reduction.  We are also grateful to Sandy Faber, Puragra
Guhathakurta and many other AEGIS members for their helpful comments
or contributions to the AEGIS database. This work is based on
observations made with the Spitzer Space Telescope, which is operated
by NASA/JPL/Caltech.  Financial support was provided by NASA through
contracts \#1255094 and \#1256790.  ALC and JAN are supported by NASA
through Hubble Fellowship grants HF-01182.01-A/HF-011065.01-A.  We
finally wish to recognize the significant cultural role that the
summit of Maunea Kea has within the Hawaiian community.

\end{document}